\documentclass[prb,preprint]{revtex4-1}

\usepackage{amsmath}  
\usepackage{amsfonts} 
\usepackage{graphicx} 
\usepackage[colorlinks=true,linkcolor=blue]{hyperref}
\usepackage{epsfig}

\begin{document}

\title{Abundance of Nanoclusters in a Molecular Beam: The Magic Numbers}

\author{Kiamars Vafayi}
\email{k.vafayi@tue.nl}
\affiliation{Department of Mathematics and Computer Science, Technische Universiteit Eindhoven, Postbus 513, 5600 MB Eindhoven, The Netherlands}

\author{Keivan Esfarjani}
\email{ke116@rci.rutgers.edu}
\affiliation{Department of Mechanical and Aerospace Engineering, Rutgers University, Piscataway, NJ 08854, USA}
\affiliation{Institute for Advanced Materials and Devices for Nanotechnology (IAMDN), Rutgers University, Piscataway, NJ 08854, USA}

\begin{abstract}
We review the theory behind abundance of experimentally observed nanoclusters produced in beams, aiming to understand their magic number behavior. It is shown how use of statistical physics, with certain assumptions, reduces the calculation of equilibrium abundance to that of partition functions of single clusters. Methods to practically calculate these partition functions are introduced. As an illustration, we compute the abundance of Lennard-Jones clusters at low temperatures, which reveals their experimentally observed magic number behavior. We then briefly review kinetic approach to the problem and comment on the interplay between chemical, mechanical and thermodynamic stability of the clusters in more generality.
\end{abstract}

\maketitle

\newpage

\section{Introduction and description of the problem} \label{int}

One can define the cluster region as the size below which the intensiveness and extensiveness of thermodynamic variables becomes invalid due to large fluctuations in these quantities. Indeed, these concepts are valid in thermodynamic limit where use of central limit theorem insures that fluctuations become infinitesimally small. In typical atomic systems, this limit is reached for the sizes which are of the order of nanometers.
In this limit, there are significant deviations in thermodynamic quantities from those predicted for the bulk matter. Namely, the surface contribution becomes increasingly important, and for even smaller sizes, due to special stability of some specific sizes, the variation with size of thermodynamic quantities becomes non-monotonic and somewhat "irregular" (which is referred to as the magic number behavior).
Thermodynamic quantities are not the only observables that behave in this way in transition from the bulk to the cluster region. In most cases, even the atomic structure of the system changes. For example, the Lennard-Jones particles \cite{kittel} form a FCC lattice in the bulk limit, but they mainly form icosahedral structures in the cluster region \cite{ike}.
Understanding the physics and the mechanisms responsible for stability in this region is important as it is the basis of 'Nanotechnology', widely applicable in designing new materials with novel properties and functionalities.

In this paper, we try to understand a statistical physics approach to treat a popular experiment which is used to synthesize nanoclusters and to investigate some properties of clusters. We aim to calculate the abundance of clusters versus their size (number of atoms) in the molecular beam produced in this experiment. We show how this problem can be reduced, under certain conditions to be mentioned later, to the calculation of the partition function (PF) of single clusters. These results are well-known, but here we try to give a pedagogical derivation and explicitly mention the underlying assumptions. Understanding of this problem has a two-fold importance;
1) the experiment (explained bellow) is one of the main methods of synthesizing nanoclusters\cite{Alonso}, and theoretical understanding of it will help to control its outcome 2) the same problem shows up in nucleation theory and in calculation of homogeneous nucleation rate, which has very important applications, e.g. the formation of water droplets in jet-engines or in clouds determining the weather conditions.

It is worth noticing that there are clusters, where the stability and magic numbers come from their electronic properties and electronic shell closure. Although the formalism is general, in this paper, we apply it to clusters of rare gas atoms, where electronic degrees of freedom are frozen, and no chemical bonding takes place. In such systems interactions are of van der Waals type and can be modeled by simple (Lenard-Jones) potentials. In this case, stability and magic numbers come from the atomic structure\cite{note1}.

We proceed later by introducing methods to calculate the PFs, and as an example, calculate low-temperature abundance of small LJ clusters which shows similar size dependent anomalous behavior (magic numbers) as is observed experimentally. Finally, we review the kinetics and its relation to mechanical and thermodynamic stability of clusters.

\section{The experimental perspective} \label{exp}

\begin{figure}[htp]
\centering
\includegraphics[width=10.5cm]{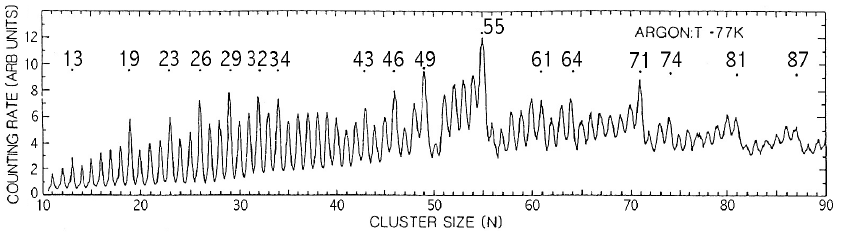}
\caption{\label{fig1} Experimental abundance vs. size of argon clusters (picture is taken from \onlinecite{harris}).}
\end{figure}

To gather the necessary assumptions for the solution, we look here at the experimental setup.
The clusters are homogeneously nucleated in a supersaturated vapor phase. This is achieved by adiabatic expansion of a rare gas into vacuum, forming a jet after passing through a capillary. Atoms strike each other in the direction of the jet flow, until their relative velocity decreases while they maintain a high translational velocity. In this way, as the gas expands in a supersonic wave, it cools down and condenses. 
For a fixed nozzle geometry, two parameters control the mean cluster size, namely the nozzle temperature $T_o$ and the stagnation pressure $P_o$. It is observed that low $T_o$ and high $P_o$ favor the condensation. 
For the purpose of detection and measurements, the jet is then usually charged by either electron bombardment, or using pre-charged atoms as nucleation centers or bombardment with laser pulses.
 It is finally detected by the time-of-flight mass spectroscopy \cite{echt,milan,harris, Alonso}. A typical output of one of these experiments from \onlinecite{harris} is shown in Figure \ref{fig1}. 
 
The way clusters are charged has an influence on the observed intensity distribution \cite{haber}; the reason for this is not yet clearly understood \cite{ike2}.
For Xe and Ar clusters, the time between the nucleation and detection is of the order of ${\rm{10}}^{{\rm{ - 5}}} {\rm{ - 10}}^{{\rm{ - 3}}} {\rm{sec}}$ (depending on the experimental setup) and the time needed for reaching  the final state is of the order of ${\rm{10}}^{{\rm{ - 8}}} {\rm{sec}}$ (depending on cooling mechanism, independent of initial conditions) \cite{soler} therefore we are faced with a system in either mainly its \emph{equilibrium state} or perhaps a special \emph{stationary state}\cite{note2}.
In an experiment with argon clusters, the initial conditions were ${\rm{T}}_{\rm{o}} {\rm{ = 77}}\;{\rm{K,}}\;{\rm{P}}_{\rm{o}} {\rm{ = 70}}\;{\rm{Torr}}$ \cite{harris}.

\section{Assumptions of the solution} \label{assump}

Based on the considerations of the previous section, we make the following assumptions to be able to simplify and solve the problem of calculating the abundance of clusters:

1) The whole system is in thermodynamic equilibrium.

2) The system is composed of several subsystems, each containing clusters with well-defined and equal sizes.

3) The only interaction between these subsystems is that they can interchange atoms, enabling the system to reach and remain in thermal equilibrium. This interaction is assumed to have no effect on the value of PFs.

4) The volume of the whole system is fixed.

5) The density is low enough so that interactions between clusters at equilibrium are negligible.

Arguably, the most controversial assumption here is the first one. Since the non-equilibrium effects (kinetics) are known to play role in formation of nanoclusters. In fact, those are the
means by which experimentalists control the growth in their samples, by modifying the experiment setup. The kinetics change the reaction paths and therefore it controls the outcome of a given experiment. However to incorporate the non-equilibrium effects is very difficult and beyond the level of this paper. In fact doing so is not possible in general and requires taking into account the details of the particular experiment setup.
In this paper we assume the equilibrium hypothesis to be approximately valid, in order to calculate the abundance of clusters in more generality. We then compare the calculations with the experimental results for argon. Moreover, this comparison enables one to decide how far is the real system from its theoretical equilibrium state. We will briefly comment on the kinetic approach to the abundance problem in the section \ref{kinetic} and mention the connection between chemical, mechanical and thermodynamic stability of the clusters in section \ref{stability}.

From assumption-1 we deduce that the internal temperature of clusters should be the same as their translational temperature and so the whole system has a single temperature ${\rm{T}}$. As a result of assumptions-1, 3 and 4, the free energy $A$
of the whole system is minimal \cite{huang}.
We do not need then to know how the system has reached the equilibrium state. The mechanism and processes that are behind the macroscopic observables and thermodynamic variables are not important as long as the state is equilibrium, since they only affect the relaxation time and relaxation path of reaching the equilibrium \cite{huang,reif}. 
Furthermore, our solution is based on considering particles as classical. This is justified when the atomic size is much larger than the thermal or the DeBroglie wavelength of the particles. In the limit when the temperature is near zero, or particles have a very small kinetic energy, atoms will experience fluctuations even at the absolute zero: the so called zero-point quantum fluctuations, which is a consequence of Heisenberg uncertainty principle \cite{huang}. 
Here, we assume that the classical criterion is satisfied. The thermal wavelength of Ar atoms at temperature $60K$
is about $0.36 \AA $ and the radius of an argon atom is ${\rm{1}}{\rm{.8{\AA}}}$
. If the classical criterion fails, quantum PFs should be used instead of classical ones but the general formalism remains unchanged. 

\section{ The Statistical Method: Equilibrium criteria and partition functions} \label{stat}

For the sake of brevity, we assume the following notations and definitions in this section and later in the paper:

\begin{tabular}{ |l | l ||| l | l |}
\hline
  $\alpha  - cluster$ & A cluster containing $\alpha$ atoms & $
S_\alpha$  & Subsystem containing $\alpha  - clusters$  \\
  $N_\alpha$ & Number of $\alpha  - clusters$  & $n_\alpha  \equiv \frac{N_\alpha}{V}$ & Density of $\alpha  - clusters$  \\
  $N$ & Total number of atoms  & $ \rho\equiv \frac{N}{V}$  & Total density   \\
$ m_0$  & Mass of atoms & $A _ \alpha$  & Free energy of $
S_\alpha$\\  
$A$ & Total free energy  & $ \mu _ \alpha$ & Chemical potential of $
S_\alpha$\\
\hline
\end{tabular} \newline
Therefore, in this notation, we have that
\begin{equation}
\label{def_a}
N = \sum\limits_\alpha  {\alpha N_\alpha  }, \; A = \sum\limits_\alpha  {A_\alpha  } , \;
\mu _\alpha   = \frac{{\partial A_\alpha  }}{{\partial N_\alpha  }}
.
\end{equation}
At equilibrium, the following thermodynamic potential is minimum:
\[{\rm{\Phi }} = A - \eta N,\]
where $\eta $
is a Lagrange multiplier, introduced to impose the constraint of constant number of particle. Afterwards, $
N_\alpha$'s can be treated as independent variables. From \ref{def_a} we obtain:
\[\delta A = \sum\limits_\alpha  {\delta A_\alpha  }  = \sum\limits_\alpha  {\frac{{\partial A_\alpha  }}{{\partial N_\alpha  }}} \delta N_\alpha   = \sum\limits_\alpha  {\mu _\alpha  } \delta N_\alpha 
 \;\; \rm{and} \;\; \delta N = \sum\limits_\alpha  {\alpha \;\delta N_\alpha  .}  \]
Now to find the equilibrium condition, the variation of ${\rm{\Phi }}$
equal is set to zero:

\[
\delta {\rm{\Phi  = }}\sum\limits_\alpha  {(\mu _\alpha   - \eta \alpha )} \delta N_\alpha  {\rm{ = 0}}
.\]
Since all the $\delta N_\alpha$'s are independent, we obtain our main equilibrium criterion:
\begin{equation}
 \label{equil_crit}
 \eta := \frac{{\mu _\alpha  }}{\alpha }   = Constant.
\end{equation}
To calculate the free energy $A_\alpha  $, the PF of the subsystem $S_\alpha$
(which contains only $\alpha  - clusters$) is needed. Suppose that the PF of one $\alpha  - cluster$
 in volume $V$ is $Z_\alpha ^{(1)}$. Our system is composed of $N_\alpha$
of these clusters, and the interaction between these clusters is negligible (assumption 5) so the subsystem $S_\alpha $ is composed of $N_\alpha$ independent and indistinguishable $\alpha  - clusters$, therefore its PF is
\[
Z_\alpha   = \frac{1}{{N_\alpha  !}}\left( {Z_\alpha ^{(1)} } \right)^{N_\alpha  } 
.\]	
The classical PF of one $\alpha  - cluster$  is
\[
Z_\alpha ^{(1)}  = \frac{1}{{\alpha !h^{3\alpha } }}\int {e^{ - \beta H} d^{3\alpha } p} \;d^{3\alpha } q
,\]
where $h$ is the Planck constant, and $p$
and $q$ are the momentum and position variables respectively. The total energy is \[
H = E_K  + E_P 
,\]
where the kinetic energy $E_K$ depends only on $p$, and the potential energy $E_P$  is assumed to depend only on $q$. Therefore
\[
Z_\alpha ^{(1)}  = \frac{1}{{\alpha !}}\left( {\frac{1}{{h^{3\alpha } }}\int {e^{ - \beta E_K } d^{3\alpha } p} } \right)\left( {\int {e^{ - \beta E_P } } d^{3\alpha } q} \right)
.\]
All possible states of the system with respect to $p$
will be counted once and only once if we count all values of the momentum vector of every atom within the cluster. The integration over $p$ is therefore similar to that of an ideal gas with $\alpha$
 particles \cite{huang};
\[
Z_\alpha ^{(1)}  = \frac{1}{{\alpha !}}\left( {\frac{1}{{\lambda ^3 }}} \right)^\alpha  \int {e^{ - \beta E_P } d^{3\alpha } q} 
\]
where $\lambda$ is the thermal wavelength:
\[\lambda  := \left( {\frac{{h^2 }}{{2\pi m_0 k_B T}}} \right)^{1/2} .\]
To count the states of the position part, two steps are followed. First we count all the configurations of the particles in the center of mass coordinate system. Second, we move the center of mass through all of the volume of the container.
As a result, the position part of the PF can be written as
\[
\int {e^{ - \beta E_P } d^{3\alpha } q}  = V\Lambda _\alpha  
,\]	
where $\Lambda _\alpha$ is integration over only one $\alpha  - cluster$
 in the center of mass coordinate system and $V$ is the volume of the container. We have
\[
\Lambda _\alpha   = \int {e^{ - \beta \sum\limits_{i < j} {V(\overrightarrow {r_{ij} } )} } } d^{3(\alpha  - 1)} q
,\]	
where $
V(\overrightarrow {r_{ij} } )$
 is the assumed pair potential between atoms in the cluster and is a function of the distance vector between each pair of atoms $\overrightarrow {r_{ij} }$. This integral has $3(\alpha  - 1)$
 variables due to the extra condition:
\[
\sum\limits_{j = 1}^\alpha  {\overrightarrow {r_j } }  = 0
.\]	
Finally we get
\begin{equation}
 \label{PF}
 Z_\alpha ^{(1)}  = \frac{V}{{\lambda ^{3\alpha } }}\frac{{\Lambda _\alpha  }}{{\alpha !}}.
\end{equation}

\section{ Density of $\alpha -clusters$}\label{dens}

Now we are ready to combine the equilibrium criteria and the PFs to obtain the density of $\alpha  - clusters$.
By definition, the free energy is:
\begin{equation}
\label{freeE}
 A_\alpha   =  - K_B T\;Log\,Z_\alpha ,
\end{equation}
and therefore,
\[
Log\;Z_\alpha   =  - Log(N_\alpha  !\;) + N_\alpha  Log\left( {\frac{V}{{\lambda ^{3\alpha } }}\frac{{\Lambda _\alpha  }}{{\alpha !}}} \right)
.\]
Together with \ref{cuttoff}, these give rise to
\[
\mu _\alpha   = \frac{{\partial A_\alpha  }}{{\partial N_\alpha  }} = K_B T\left[ {Log\,N_\alpha   - Log\left( {\frac{V}{{\lambda ^{3\alpha } }}\frac{{\Lambda _\alpha  }}{{\alpha !}}} \right)} \right]
.\]
From the equilibrium criterion \ref{equil_crit} we deduce:
\begin{equation}
\label{ndensity}
n_\alpha   = \left( {\frac{{e^{\beta \eta } }}{{\lambda ^3 }}} \right)^\alpha  \frac{{\Lambda _\alpha  }}{{\alpha !}}
.
\end{equation}
In principle, $\eta$ can be found from the constraint
\[\sum\limits_\alpha  {\alpha N_\alpha  }  = N ,\]
 or      \begin{equation} \label{etaconstrait}
\sum\limits_{\alpha  = 1}^\infty  {\alpha n_\alpha  }  = {\raise0.7ex\hbox{$\rho $} \!\mathord{\left/
 {\vphantom {\rho  {m_0 }}}\right.\kern-\nulldelimiterspace}
\!\lower0.7ex\hbox{${m_0 }$}}
\end{equation}
We can also write the density in another identical form. From free energy of one $\alpha  - cluster$ \ref{freeE} together with \ref{PF} we obtain:
\[
Z_\alpha ^{(1)}  = \frac{V}{{\lambda ^{3\alpha } }}\frac{{\Lambda _\alpha  }}{{\alpha !}} = e^{ - \beta A_\alpha ^{(1)} } 
,\]	
which in conjunction with \ref{ndensity} gives:
\[
n_\alpha   = \frac{{\left( {e^{\beta \eta } } \right)^\alpha  }}{V}e^{ - \beta A_\alpha ^{(1)} } 
.\]
In experiments, $n_\alpha $
is not directly measured; instead, $I_\alpha  $
 which is the intensity of the $\alpha  - cluster$
 beam, is measured. As the observed intensity is proportional to the cluster current, it is reasonable to assume that $
I_\alpha   \propto {\rm{v}}_\alpha  n_\alpha  
$
, where $
{\rm{v}}_\alpha  
$
is the average speed of $\alpha  - clusters$
 and is proportional to $
{\raise0.7ex\hbox{${\rm{1}}$} \!\mathord{\left/
 {\vphantom {{\rm{1}} {\sqrt {\rm{\alpha }} }}}\right.\kern-\nulldelimiterspace}
\!\lower0.7ex\hbox{${\sqrt {\rm{\alpha }} }$}}
$
, therefore, the observed intensity should be
\begin{equation}
\label{final_abundance}
 I_\alpha   = \frac{\chi }{{\sqrt \alpha  }}n_\alpha   = \frac{\chi }{{V\sqrt \alpha  }}\,\,e^{ - \beta (A_\alpha ^{(1)}  - \alpha \eta )} 
,
\end{equation}
where the quantity $\chi$ is a proportionality constant.

\subsection*{\emph{Intuitive meaning of $\eta$}}
We can think of $\eta$ as a \emph{generalized} chemical potential obtained through \ref{etaconstrait} as a function of $\rho$. On the other hand, for every value of $\eta$, the equation \ref{etaconstrait} defines a corresponding density $\rho(\eta)$. The bigger the $\eta$ the bigger $n_\alpha$, and the slower it will decay with increasing $\alpha$. Therefore $\rho(\eta)$ is an increasing function of $\eta$. Thus $\eta$ is like a parameter that tunes the system total density, it can also be related to the gas-liquid phase transition. In fact, there is a critical value of $\eta_c$ above which $\rho(\eta)$ becomes infinite (physically it means the system condenses fully into liquid phase). To see this, note that for large clusters (as in thermodynamic limit) $A_\alpha=K\alpha$, where $K$ is a constant independent of $\alpha$. The sum in \ref{etaconstrait} will obviously diverge if $\eta>K$. 

Intuitively, very small $\eta$ corresponds to a vapor with smaller size clusters and the vice versa. The region which is more interesting, however, is somewhere in between, where $\eta$ is comparable to $K$, and as a result clusters of widely varying sizes can coexist.

\section{ The cutoff volume of integration }\label{cutt}

The integral in $\Lambda _\alpha $ should be integrated only over a small volume relative to $V$, since not all the volume of the container belongs to one cluster. As a good approximation we can say that
\[ \nu  \equiv {V \mathord{\left/
 {\vphantom {V {\sum\limits_{\beta  = 1}^\infty  {\beta N_\beta  } }}} \right.
 \kern-\nulldelimiterspace} {\sum\limits_{\beta  = 1}^\infty  {\beta N_\beta  } }} = \frac{V}{N}\]
belongs to every atom and similarly
\[\nu _\alpha   = \alpha \,\nu  = {{\alpha \,V} \mathord{\left/
 {\vphantom {{\alpha \,V} {\sum\limits_{\beta  = 1}^\infty  {\beta N_\beta  } }}} \right.
 \kern-\nulldelimiterspace} {\sum\limits_{\beta  = 1}^\infty  {\beta N_\beta  } }}\]
belongs to every $\alpha  - cluster$.
At temperatures where the clusters form, only a small portion of the volume of integration contributes to $\Lambda _\alpha$
 significantly. It means that if the volume of integration already includes this portion, then changing the volume further has no appreciable effect on the value of the integral, that is
 \begin{equation}
 \label{cuttoff}
  \frac{{\partial \Lambda _\alpha  }}{{\partial N_\alpha  }} \simeq \frac{{\partial \Lambda _\alpha  }}{{\partial \nu _\alpha  }} \simeq 0.
 \end{equation}
In conclusion, when we use the proper order of magnitude for $\nu$
 the results will not depend strongly on the exact value of $\nu$, particularly at lower temperatures\cite{lee}.

\section{Discussion of the statistical physics approach }\label{adv}

\begin{figure}[htp]
\centering
\includegraphics[width=10.5cm]{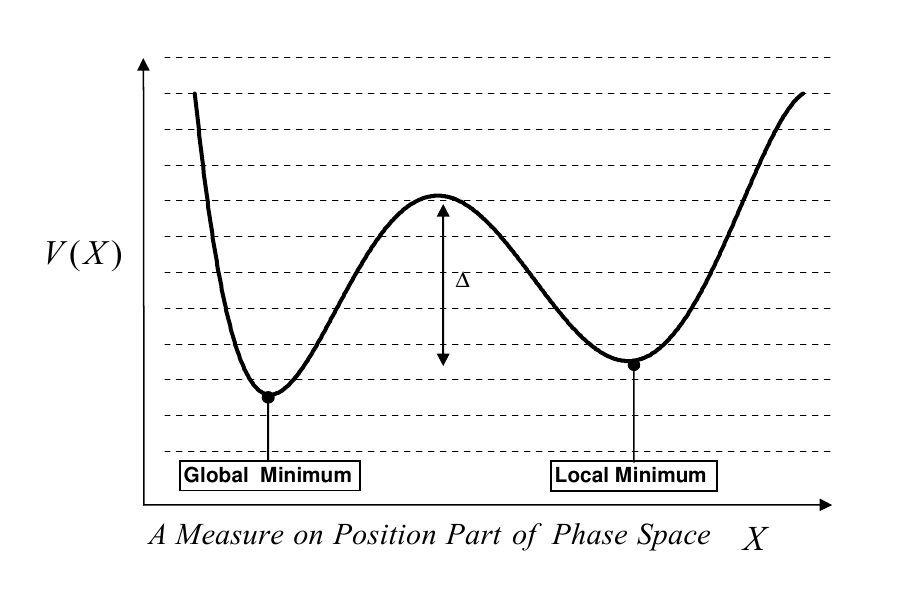}
\caption{\label{fig2} Potential energy of two isomers of one $\alpha  - cluster$}
\end{figure}

Comparing this statistical physics approach to some other methods of solving this problem \cite{harris,ike2,ike3} can deepen our understanding of this method. Here are three remarks about this method:

1) In principle, we should take into account every possible configuration of the atoms in a cluster. The only thing measured in the experiment is the number of atoms in the cluster and not their configuration. Indeed, what one observes in the experiment as $\alpha - clusters$ belong to an ensemble of clusters with $\alpha $ atoms but with different configurations. However, the low temperature condition makes some special configurations (or isomers) of one $\alpha  - cluster$ more abundant.
To illustrate this point suppose that in another method one treats the problem on the sole basis of the global minimum energy configuration of the cluster. However, it is probable, and sometimes this is the case, that there are other isomers with energies very close to the global minimum and there are large barriers of transition between these configurations in the phase space. For example, in Figure \ref{fig2}, if $\Delta  \gg k_B T$
it is not expected that a dynamical simulation of the cluster at temperature $T$
will result in a transition from local minimum to global minimum or vice versa in a reasonable time.

However if energies of global and local minimum are comparable with respect to $k_B T$
then the PF will contain significant contribution from the local as well as the global minimum.
As an example, a 38 atoms Lennard-Jones cluster has an energy of ${\rm{ - 173}}{\rm{.928427}}\varepsilon $
 in its global minimum structure whereas another local minimum structure has an energy of ${\rm{ - 173}}{\rm{.252378}}\varepsilon $
 in units of $\varepsilon$
 in the Lennard-Jones potential \cite{kittel}. While they differ only by 0.4\%, a barrier between these two structures can result in a complete neglect of one of them. 
Note, on the other hand, that a disagreement between theory and experiment can still occur if at high temperatures the local minimum isomer has lower free energy, and thus, not having enough kinetic energy to cross the large 
potential barrier, then the system may remain, after the annealing process, in this minimum and one would not find {\emph {experimentally}} the true global minimum. This problem can only be overcome if the annealing rate is very slow.

2) In some other methods \cite{ike2,ike3} the process of relaxation of the total system from non-equilibrium  states, which is very compute-intensive, is simulated. However, according to the statistical physics method, doing so is unnecessary in the equilibrium setting.

3) This solution is based on a few plausible assumptions, so the failure in its application can be readily traced back to these assumptions or the way the PFs have been computed.

\section{ Methods for evaluation of the partition function }\label{method}

At first it may seem that the problem is solved thoroughly; we just need to compute the integral of $\Lambda _\alpha$. However, even for a middle-sized cluster of 31 atoms, $\Lambda _\alpha$ is a 90 dimensional integral and is therefore compute-intensive. If one were to divide each dimension of the integral to 100 equal parts to estimate the integral by a simple sum, the total number of points in the phase space at which we ought to calculate the potential energy would be ${\rm{10}}^{{\rm{180}}}$ which is a huge number. So the problem of integrating the PF is a challenge on its own. Following is a list of some of the most widely used methods for evaluating the PF; details of them can be found in the given references.

(i) Importance Sampling Technique: The states in the sum are sampled by the Monte Carlo method \cite{gould,negele,binder} according to an integrable distribution function similar to the Boltzmann factor itself so that the fluctuations of the integrand are small. 

(ii) Low Temperature Harmonic Approximation: Assuming the Hamiltonian to be separable, and the cluster be a solid, the PF is factored into the product of rotational, vibrational, translational, (and eventually electronic PFs), each being easily calculated numerically provided that the rotational, vibrational and electronic excitations are known 
\[
Z = Z_{vib} Z_{rot} Z_{tr} (Z_{el} ) .
\]
By changing the coordinate system to the normal modes of the coupled oscillators, we will have $3\alpha  - 6$
 independent harmonic oscillators for which the PF is known: 
\[
Z_{vib}  = \prod\limits_{{\rm{mode}}\,n} {\left( {2\sinh \,[\beta \hbar \omega _n /2]} \right)^{ - 1} } 
\] 
Likewise, $Z_{rot}$ and $Z_{tr}$ can be obtained by means of quantum rigid rotator and quantum ideal gas approximations, respectively \cite{wilson}.

(iii) The Multiple Histogram Method: The PF is written as the energy integral of the Boltzmann factor times the density of states. The latter is computed by means of computing histograms of the energy distribution at different neighboring temperatures \cite{newman,ferren,landau}. 

(iv) High Temperature Virial Expansion: For a system where the interaction of particles is weak relative to their kinetic energy, the PF can be expanded at high temperatures by using the virial expansion around the ideal gas (noninteracting) limit. At high temperatures, few of the virial coefficients suffice to get reasonable results \cite{reref_st}.

(v) Moderate Temperature Statistical Physics Methods: In these methods, the PF is determined from the knowledge of its value in some limit (e.g. high temperature virial expansion or the low temperature harmonic approximation) plus the dependence of the average potential energy on temperature which can be obtained by means of Monte Carlo or Molecular Dynamics methods \cite{lee,binder,garcia,McGinty}

\section{ Magic numbers }\label{magic}

\begin{figure}[htp]
\centering
\includegraphics[width=10.5cm]{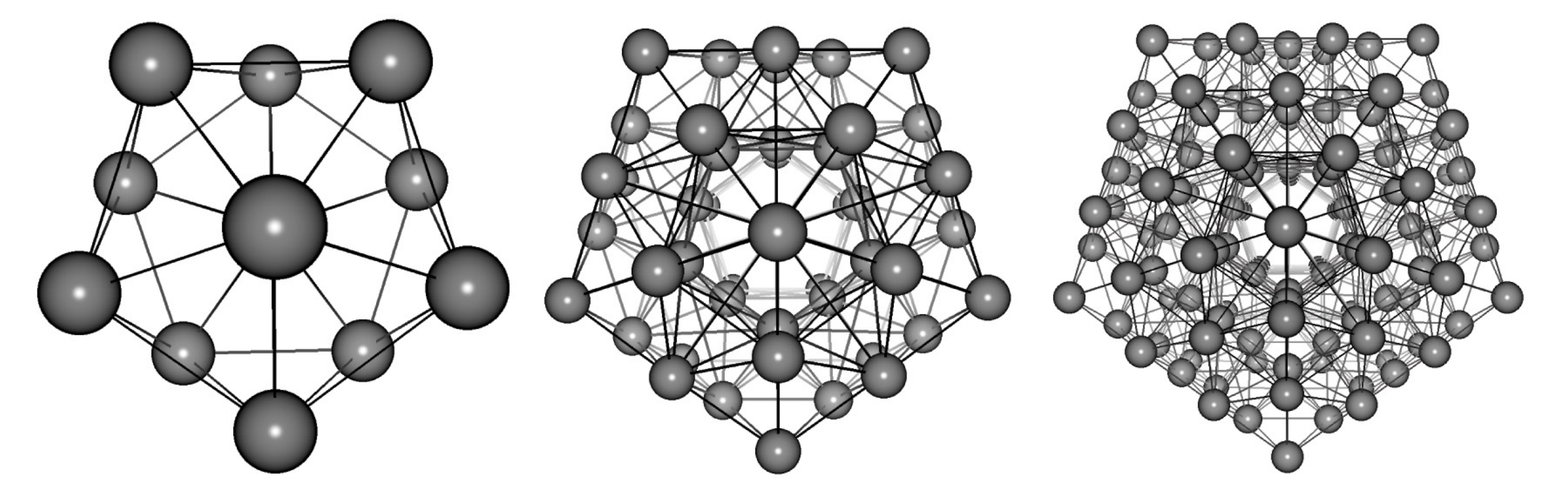}
\caption{\label{fig3} Multi-layer icosahedral structures of the magic numbers $\alpha  = 13,55,147$}
\end{figure}

An interesting problem about cluster abundance has been to understand magic numbers \cite{echt,harris,ike2,ike3}. The experiments have shown that some clusters are magic, i.e. some clusters that contain certain number of atoms are evidently more abundant than the others (figure \ref{fig1}). Magic number series for each material depend on the detail of the interaction between particles. For example, the sequence of numbers:
\[N_{magic} (n) = 1 + \sum\limits_{k = 1}^n {(10k^2  + 2)}  = 13,\;55,\;147,\;...\]
corresponding to atomic icosahedral (figure \ref{fig3}) shell closure, are one of the most important magic number series, observed experimentally in the rare gas atomic clusters \cite{ike3}. A LJ cluster whose minimum energy structure is symmetric is likely magic. In principle, a detailed computation of abundance versus size based on the statistical approach of this paper should reveal the magic numbers.
We should mention that in the case of non-metals, the magic numbers and enhanced stability comes from the geometrical arrangement of the atoms, leading often to symmetric structures like icosahedra etc..., while in the case of metallic clusters\cite{metalnano1,metalnano2,metalnano3} such as Sodium, magic numbers come from electronic shell closure, leading to a large gap between occupied and empty electronic states. In this case the ionization potential, i.e. the energy to remove an electron from the closed electronic shell becomes very large, while in the rare-gas atoms case, the energy to remove an atom from the atomically closed shell cluster is very large. 

\section{ Application to small LJ clusters }\label{lj}

\begin{figure}[htp]
\centering
\includegraphics[width=10.5cm]{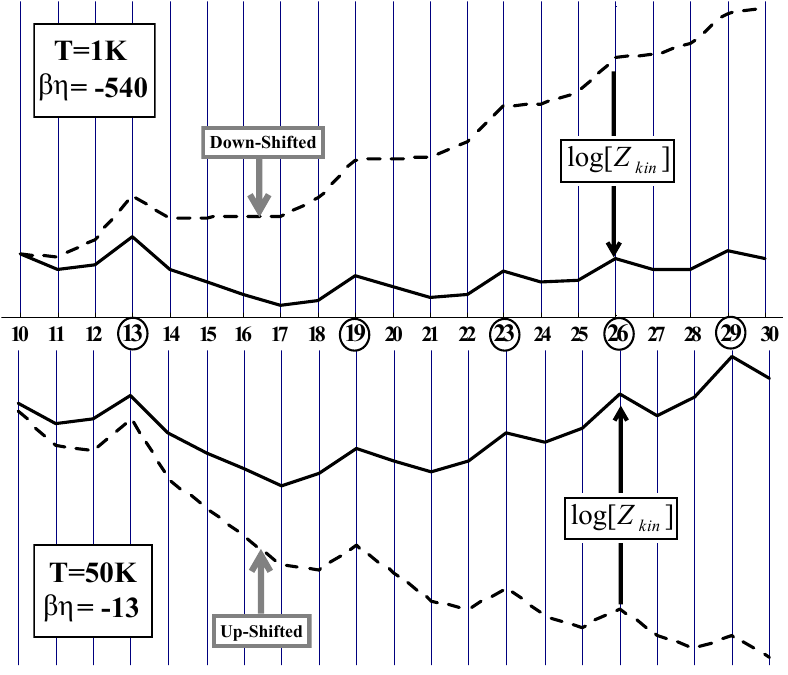}
\caption{\label{fig4} Computed abundance (arbitrary Log-Scale) vs. argon cluster size, Circles are drawn around experimental magic numbers. 
The curves are shifted vertically for a better comparison. $Z_{kin} $ is the kinetic energy contribution to the total partition function (note that the two plots are not in a same scale).}
\end{figure}

As an illustration of the method, we have calculated the relative abundance of the small Lennard-Jones clusters ranging in size from 10 to 30 within the Harmonic Approximation (HA) \cite{wilson}. The global minimum structures were extracted from \cite{camdata}. Then, in order to compute the vibrational and rotational PFs, the force constants and vibrational frequencies, as well as the three moments of inertia were calculated respectively. Electronic degrees of freedom were neglected since the temperature is very small compared to the electronic excitation energies.
For comparison, we have plotted in figure \ref{fig4} the abundance defined in equation \ref{final_abundance} versus the particle number with and without the kinetic PF contribution at T=1K and 50K, even though HA ceases to be valid at 50 K. One can notice that the main magic features come from the potential energies and finite temperature, and kinetic effects influence only the relative abundance. It is remarkable that all the magic numbers in this range are in accordance with the experimental results in figure \ref{fig1}.  

However, the relative magnitudes of abundances do not agree with experiments; relative abundances of magic numbers are too high. 
In the higher temperature (50K) this sharpness of magic peaks has been greatly softened, but has not been fully resolved. 
In fact, the real experiments are done at even a higher temperature $~70K >> 1K$ where the HA is no longer valid and the clusters are not solid but are more liquid-like.
 It turns out that for the considered Lennard-Jones potential and temperatures of interest, the HA underestimates the real potential energy and leads 
to values for the PF which are much higher than the true PF. 

\section{Kinetic approach}\label{kinetic}
Here we briefly review the kinetic approach to the cluster abundance problem. Suppose that the system is composed of a total of $M$ atoms and they are grouped in clusters such that there are $N_{\alpha}$ of $\alpha-cluster$s. Therefore the state of the system can be represented with a vector $(N_1,...,N_{M})$. By $N_1$ we mean the number of single atoms and its obvious that $N_{M}$ can be maximum equal to one\cite{note3}.

The kinetic approach is concerned with the rate of collisions (i.e. the number of collisions per second) and the rate of disintegration of the clusters in such a vapor-phase system. In principle there can be any kind of collision involving two or more clusters of sizes $1,...,M$ with various possible outcomes. At relatively low densities, one can reasonably argue that two cluster collisions are the most common type of collisions, since the probability of three (or more) cluster collision is much less, and the corresponding average time between two such collisions is much larger. Therefore considering only two cluster collisions captures the main aspects of kinetic phenomena. The outcome of such collision could be the merging of the two cluster into one bigger cluster, or the emergence of two or more new clusters of different type. To keep the calculations simpler its usually assumed that the collisions only result in merging. Similarly its assumed that the most common disintegration process is that of one cluster converting into two clusters of smaller size but with a same total number of atoms\cite{note4}.

Assuming that the system has the Markov property, i.e. that the transition probabilities depend only on the current state of the system and not on its history, one can write the following set of `master equations' for the time evolution of the cluster numbers
\begin{equation}
\frac{d N_i}{dt}= \sum_j N_j P_{j\to i} - N_i \sum_j P_{i\to j}
\end{equation}
for each cluster size $i$. Here $P_{j\to i}$ is the {\emph{mean total rate}} at which a $j-cluster$ converts to an $i-cluster$. Although this equation seems simple at first look, the main problem is to be able to calculate these rates. For instance, note that $P_{j\to i}$ depends, in a non-trivial way, certainly on $N_j$ and also the rest of $(N_1,...,N_{M})$, since the transition $j\to i$ can be composed of collision or disintegration involving other clusters, making the problem very complicated. So to be able to proceed one needs further simplifactions. In the framework of Classical Nucleation Theory (CNT) \cite{Kalikmanov, abrahamn, Ford} it is a common approximation that the transitions in the system are only due to interaction of clusters with single atoms, i.e. two possible mechanisms:

{\bf 1)} A single atom collides with an $i-cluster$, merging to create an \newline 
$(i+1)-cluster$. As a result $N_1\to N_1-1$, $N_i\to N_i-1$ and $N_{i+1}\to N_{i+1}+1$.

{\bf 2)} An $i-cluster$ disintegrates by evaporating a single atom and turning into an $(i-1)-cluster$. As a result $N_1\to N_1+1$, $N_i\to N_i-1$ and $N_{i-1}\to N_{i-1}+1$. 

Assuming that these two processes occur with rates $C_i$ and $E_i$ respectively, its usual to write the following set of master equation for the system,
\begin{equation}
\frac{d N_i}{dt}= I_{i-1} - I_i
\end{equation}
in which $I_i:=C_i N_i - E_{i+1} N_{i+1}$. These are the so called the Becker-Doring equations \cite{BDeq}.
One is usually interested in the steady-state non-equilibrium solution of the master equation, i.e. the state in which the (average) number of clusters is not changing with time thus satisfying
\begin{equation}
\frac{d N_i}{dt}=0.
\end{equation}
Or the equilibrium solution that should additionally satisfy the condition of detailed balance \cite{reif};
\begin{equation}
I_i=0 \quad or \quad C_i N_i = E_{i+1} N_{i+1}.
\end{equation}
These solutions should in principle provides us with the abundance of the clusters in the given conditions.
Although after several approximations the resulting Becker-Doring equations are much simplified, the accurate calculation of the rates $C_i$ and $E_i$ is highly non-trivial, specially for smaller size clusters. Within the CNT one usually suppose further simplifying assumptions to be able to estimate $C_i$ and $E_i$ or at least their approximate dependence on the size of the cluster $i$ and temperature and pressure. Usually it is assumed that the clusters have a well defined and smooth spherical shape with a radius determined by $i$. As a result, its not surprising that calculations of abundance based on this method does not show any size dependent anomalous behavior (magic numbers) \cite{soler, soler2}. We believe the reason for this to be the series of approximations mentioned and the fact that these calculation do not take into account the fine structural and energy differences between small size clusters which are crucial in determining their magic numbers. Nevertheless, such kinetic approaches have been useful to understand the average behavior of non-equilibrium vapors in an at least qualitative fashion \cite{Kalikmanov}.

\section{ Other criteria for stability }\label{stability}
Our solution to the cluster abundance in the section \ref{stat} was based on the energetics of the clusters and their free energy. As we mentioned in the section \ref{assump}, in an actual experiment, the abundance might be affected by other factors such as the kinetics of the reactions leading to the final outcome \cite{Alonso}. Here, we would like to point out that the stability of a cluster or molecule depends on other factors as well. One is of course the total energy ({\bf the thermodynamic stability}), but one can also define {\bf kinetic} and {\bf mechanical stability} as well. These could be related as we will show below.

Kinetic or chemical stability means that the cluster will not or hardly react to reactants or neighboring molecules. They do not tend to deform or form chemical bonds. A simple example are the rare gas atoms which have their electronic shells fully filled. In other words, their ionization energy will be very large. In the context of molecules and clusters, a filed electronic shell corresponds to a large HOMO-LUMO gap \cite{Levine,Szabo}. Systems with a large electronic gap do not tend to hybridize as there is hardly any energy gain in doing such a thing. On the other hand, a small or zero-gap system, which is also called an open-shell system, can easily share electrons with a neighboring system and gain energy from hybridization. Metallic clusters\cite{metalnano1,metalnano2,metalnano3} are a good example of kinetically unstable systems. 

Mechanical stability, as its name indicates, has to do with how much the total energy will increase if the system is deformed. Basically, if the system is already relaxed to its ground state, the potential energy would be positive definite, meaning that its Taylor expansion in powers of atomic displacements, up to second order would remain positive for any arbitrary (small) displacement. When the latter quadratic form is diagonalized, the vibrational modes become decoupled into "normal modes" with positive stiffness, written as $k_{\lambda}=m \omega_{\lambda}^2 $, $\lambda$ labeling the normal mode (the potential energy would then be $ \sum_{\lambda} m \omega_{\lambda}^2 x_{\lambda}^2/2$). The lowest normal mode frequency indicates how soft the system is. Systems in which the lowest normal mode frequency is "large" are mechanically stiff, and do not want to deform, while soft clusters have their lowest $k_{\lambda} $ small but still positive. A negative stiffness implies the cluster is mechanically unstable and would further gain potential energy under appropriate deformation. 

It can easily be imagined that an open-shell system would most likely be mechanically soft, if the energy gain, when put in contact with another system, is achieved through atomic deformation. In other words, through chemical hybridization, an energy gap could be opened, leading to energy gain. If this is accompanied by a deformation, then one could state that mechanical stability can be correlated with kinetic or chemical stability. To explain this better, we refer to the concept of "deformation potential". It is the proportionality between the shift in electronic energy levels and the atomic displacements, if the latter are small. For a system with small or zero gap, specific atomic displacements which make the HOMO and LUMO levels switch places would not cost much energy. This leads to a soft system. In other words, small gap systems are likely to also be mechanically soft, although this is not a general rule.   

These two concepts are not, however, necessarily related to thermodynamic stability. For both of them, the reference configuration only needs to be a {\it local} energy minimum, also called a metastable state. As far as we can see, there is no relationship between the HOMO-LUMO gap of the lowest energy state and that of another metastable state. One could argue that if a cluster has a large gap, it is likely to be in the most stable configuration, although this is not a general rule and exceptions to this have been seen. Likewise there is no relation between the energy of a cluster and its mechanical softness. The ground state could be mechanically softer than a metastable isomer.

\section{ Conclusion }\label{conc}

We have discussed the problem of cluster abundance in a simple language. It has been shown that using basic equilibrium statistical mechanics, one can reduce the problem to a calculation of single cluster PFs. Some standard methods to compute PFs were mentioned. Finally we have illustrated the application of this method to calculate the abundance and reveal magic numbers in small LJ clusters using the harmonic approximation which is valid at low-temperatures. We briefly reviewed the kinetic approach, mentioned the difficulties it has to calculate the magic numbers and commented on the connection between chemical, mechanical and thermodynamic stability of the clusters.
Straightforward extensions of this paper could be calculation of the abundance of charged clusters, i.e. including the Coulomb's potential, or extensions to other substances and other empirical potentials, and possible comparisons with relevant experiments. It can be instructive to calculate the PFs based on various methods mentioned in \ref{method} and to see how do they compare.


\begin{thebibliography}{99}

	\bibitem{kittel} C. Kittel, Introduction to Solid State Physics, 7th edition (John Wiley \& Sons, 1996), Chap. 3

	\bibitem{ike} T. Ikeshoji, G. Torchet, M.-F. de Feraudy, and K. Koga, Phys. Rev. E 63, 031101 (2001).

	\bibitem{Alonso} Julio A. Alonso, Structure and Properties of Atomic Nanoclusters (Imperial College Press, 2005)

	\bibitem{note1} {One can say that in the case of rare gasses too, the Lenard-Jones interaction potential is a consequence of van der Waals forces and thus of the electronic structure of the atoms. While this is true, the fact is that the force between rare gas atoms can be modeled rather accurately by the Lenard-Jones potential, which is an empirical potential and significantly simpler than trying to do calculations based on the electronic structure and using quantum mechanics.}

	\bibitem{echt} O. Echt, K. Sattler, and E. Recknagel, Phys. Rev. Lett.47, 1121 (1981).

	\bibitem{milan} P. Milani and S. Iannotta, Cluster Beam Synthesis of Nanostructured Materials (Springer, 1999), Chap. 4, p.96.

	\bibitem{harris} I. A. Harris, R. S. Kidwell, and J. A. Northby, Phys. Rev. Lett. 53, 2390 (1984).

	\bibitem{haber} H. Haberland, Clusters of Atoms and Molecules (Springer, 1994), p.374.

	\bibitem{ike2} T. Ikeshoji, B. Hafskjold, Y. Hashi, and Y. Kawazoe, Phys. Rev. Lett. 76, 1792 (1996).

	\bibitem{soler} J. M. Soler and N. Garcia, Phys. Rev. A 27, 3307 (1983).

	\bibitem{note2} {A stationary state is a state that does not change with time, however it is not necessarily an equilibrium state.}

	\bibitem{huang} K. Huang, Statistical Mechanics, 2nd edition (John Wiley \& Sons, 1987).

	\bibitem{reif} F. Reif, Fundamentals of Statistical and Thermal Physics, (McGraw-Hill, 1965).

	\bibitem{lee} J. K. Lee, J. A. Barker, and F. Abraham,  J. Chem. Phys. 58, 3166 (1973).

	\bibitem{ike3} T. Ikeshoji in Clusters and Nanomaterials (Theory and Experiment), Y. Kawazoe, T. Kondow, and K. Ohno (Eds), (Springer, 2002), Chap. 11

	\bibitem{gould} H. Gould and J. Tobochnik, An Introduction to Computer Simulation Methods, 2nd edition (Addison-Wesley, 1996), Chap. 11.

	\bibitem{negele} J. W. Negele and H. Orland, Quantum Many-Particle Systems (Addison-Wesley, 1988), Chap. 8, p.403.

	\bibitem{binder} K. Binder and D.W. Heerman, Monte Carlo Methods in Statistical Physics, 4th edition (Springer, 2002).

	\bibitem{wilson} E. B. Wilson, Jr., J.C. Decius, and C. Cross, Molecular Vibrations (Dover, 1955), Chap. 2.

	\bibitem{newman} M. E. J. Newman and G. T. Barkema, Monte Carlo Methods in Statistical Physics,  (Oxford, Clarendon Press, 1999), Chap. 8.

	\bibitem{ferren} A. M. Ferrenberg and R. H. Swendsen, Phys. Rev. Lett. 63, 1195 (1989).

	\bibitem{landau} Fugao Wang and D. P. Landau, Phys. Rev. Lett. 10, 2050 (2001).

	\bibitem{reref_st} Chapter 10 of Ref. 9 and Chapter 10 of Ref. 10.

	\bibitem{garcia} N. Garcia Garcia and J. M. Soler Torroja, Phys. Rev. Lett. 47, 186 (1981).

	\bibitem{McGinty} David J. McGinty, Vapor Phase Homogeneous Nucleation and the Thermodynamic Properties of Small Clusters of Argon Atoms, J. Chem. Phys. 55, 580 (1971)

	\bibitem{metalnano1} Daniel L. Fedlheim, Colby A. Foss, Metal Nanoparticles: Synthesis Characterization and Applications (CRC Press 2001)

	\bibitem{metalnano2} Part I of Clusters and Nanomaterials (Theory and Experiment), Y. Kawazoe, T. Kondow, and K. Ohno (Eds), (Springer, 2002)

	\bibitem{metalnano3} Roy L. Johnston, Atomic and Molecular Clusters (Taylor and Francis, 2002)

	\bibitem{camdata} The Cambridge Cluster Database, D. J. Wales, J. P. K. Doye, A. Dullweber, M. P. Hodges, F. Y. Naumkin F. Calvo, J. Hernández-Rojas and T. F. Middleton, URL http://www-wales.ch.cam.ac.uk/CCD.html.

	\bibitem{note3} {Therefore we do not distinguish between various possible configuration of a certain size clusters here, the remaining discussion is then about the average behavior of all possible configurations (isomers). However one still needs to 'define' what a cluster is. One definition is that a cluster is collection of atoms (or monomers) such that there is no atom in it that is further than a certain distance $r_c$ from the rest of the atoms in the cluster. The $r_c$ is determined based on the interaction potential of the atoms, such that when two atoms are closer than $r_c$, they still interact appreciably. See \onlinecite{Ford} for more discussion.}

	\bibitem{note4} {There is a subtle point that, after all, in this way one is indirectly taking into account the possibility of two cluster collisions that result in two new clusters (i.e. not only merging after collision). Since such processes can be imagined, as first a merging after collision, and then a disintegration into two new clusters.}

	\bibitem{Kalikmanov} V. I. Kalikmanov, Nucleation Theory (Springer, 2013)

	\bibitem{abrahamn} Farid F. Abraham, Homogeneous nucleation theory: the pretransition theory of vapor condensation (Academic Press, 1974)

	\bibitem{Ford} I.J. Ford, Statistical mechanics of nucleation: a review , Proc. Instn Mech. Engrs 218 Part C: J. Mech. Eng. Sci. (2004) 883-899.

	\bibitem{BDeq} Becker, R. and Doring, W. Kinetische Behandlung der
Keinbildung in ubersattigten Dampfen. Ann. Physik
(Leipzig), 1935, 24, 719.

	\bibitem{soler2} J. M. Soler and N. Garcia, Phys. Rev. A 27, 3300 (1983).

	\bibitem{Levine} Ira N. Levine, Quantum Chemistry (7th Edition) (Prentice
 Hall 2013)

	\bibitem{Szabo} Attila Szabo and Neil S.
 Ostlund, Modern Quantum Chemistry: Introduction to Advanced
 Electronic Structure Theory (Dover Books on Chemistry 1996)
\end{thebibliography}
\end{document}